# Flexible, solid electrolyte-based lithium battery composed of LiFePO$_4$ cathode and Li$_4$Ti$_5$O$_{10}$ anode for applications in smart textiles

Yang Liu,*[a] Stephan Gorgutsa[a] Clara Santato[a] and Maksim Skorobogatiy[a]

Here we report fabrication of flexible and stretchable battery composed of strain free LiFePO$_4$ cathode, Li$_4$Ti$_5$O$_{10}$ anode and a solid poly ethylene oxide (PEO) electrolyte as a separator layer. The battery is developed in a view of smart textile applications. Featuring solid thermoplastic electrolyte as a key enabling element this battery is potentially extrudable or drawable into fibers or thin stripes which are directly compatible with the weaving process used in smart textile fabrication. The paper first details the choice of materials, fabrication and characterisation of electrodes and a separator layer. Then the battery is assembled and characterised, and finally, a large battery sample made of several long strips is woven into a textile, connectorized with conductive threads, and characterised. Within this paper, there are two practical aspects of battery design that we have investigated in details: first is making composites of cathode/anode material with optimized ratio of conducting carbon and polymer binder material, and second is battery performance including cycling, reversibility, and compatibility of the cathode/anode materials. Finally, when casting electrodes and separator layer we mostly focused on using aqueous solutions instead of organic solvents in order to make the fabrication process environmentally friendly.

## Introduction

With the rapid development of micro and nanotechnologies and driven by the need to increase the value of conventional textile products, fundamental and applied research into smart textiles has recently flourished. Generally speaking, textiles are defined as "smart" if they can respond to the environmental stimulus, such as mechanical, thermal, chemical, electrical, and magnetic. Many applications of "smart" textiles stem from the combination of textiles and electronics (e-textiles). Most of the "smart" functionalities in the early prototypes of e-textiles were enabled by integrating conventional rigid electronic devices into a textile matrix. The fundamental incompatibility of the rigid electronic components and a soft textile matrix create a significant barrier for spreading of this technology into wearables. This problem motivated many recent efforts into the development of soft electronics for truly wearable smart textile. This implies that the electronic device must be energy efficient to limit the size of the battery used to power it. Needless to say that to drive all the electronics in a smart textile one needs an efficient, lightweight and flexible battery source. Ideally, such a source will be directly in the form of a fiber that can be naturally integrated into smart textile during weaving.

Broadly speaking, the advancements in flexible batteries have been in the following categories: (a) flexible organic conducting polymers [1-4], (b) bendable fuel cells [5], (c) polymer solar cells [6-8] and (d) flexible lithium polymer batteries [9-12]. Recently, a rechargeable textile battery was created by Bhattacharya et al. [13]. It was fabricated on a textile substrate by applying a conductive polymeric coating directly over interwoven conductive yarns. Approaches to produce stretchable and foldable integrated circuits have also been reported. This includes integrating inorganic electronic materials with ultrathin plastic and elastomeric substrates [14] and printing high viscous conductive inks onto nonwoven fabrics [15]. Stretchable, porous, and conductive textiles have been manufactured by a simple "dipping and drying" process using a single-walled carbon nanotube (SWNT) ink and the nanocomposite paper, engineered to function as both a lithium-ion battery and a supercapacitor, which can provide a long and steady power output [16, 17]. Among those flexible batteries, the lithium polymer battery has taken much attention for its potential in electric vehicle applications. It employs a solid polymer electrolyte, which can act both as the electrolyte and the separator, with the aim of improving battery design, reliability, safety, and flexibility.

There are two features shared by the majority of existing flexible batteries that make them ill-suited for applications in smart textiles. The first one is the realisation that conventional polymer electrolytes and binders used in lithium batteries to blend anode, cathode and conducting materials are processed with organic solvents, which are poisonous and caustic and, thus, do not fit well with wearables. The second one is the fact that, at present, flexible film batteries are not extrudable or drawable to form fibers or stripes, which are the necessary building block for smart textile fabrication. In this paper we report on the two improvements that we have achieved towards fabrication of a flexible, extrudable, and environmentally safe battery for smart textiles. The first one involves processing of both electrode binders and polymer electrolytes with aqueous solution rather than with organic solvents. This leads to an environmentally friendly process for the electrode and polymer electrolyte fabrication. The second improvement is the extensive use of a thermoplastic solid electrolyte both in the electrodes and a separator layer. This allows, in principle, fabrication of a battery preform that can be then drawn into a battery fiber.

In parallel with our previous research on flexible analogue electronics in fiber form (see for example capacitor fibers in [18, 19]), this paper study the possibility of finding a materials system for the design of a drawable lithium polymer battery with a view of eventually obtaining a battery-on-fiber. The cathode material used here is $LiFePO_4$. As detailed in [20], the discharge potential of $LiFePO_4$ is ~3.4 V vs. $Li/Li^+$ and no obvious capacity fading is observed for this material even after several hundred cycles. The specific capacity of $LiFePO_4$ is ~170 mAh/g, which is higher than for that of a conventional $LiCoO_2$. $LiFePO_4$ is, in fact, the first cathode material in Li batteries with low cost and abundant elements which is also environmentally benign. Due to the $LiFePO_4$ low intrinsic electronic conductivity ($10^{-9}$ $S/cm^2$), carbon-based materials are often coated on its surface; alternatively, transition elements, such as niobium, are introduced as dopants in order to improve the conductivity of $LiFePO_4$ by 4-8 orders of magnitude [21, 22]. The olivine structure of $LiFePO_4$ and the remaining phase $FePO_4$ after the lithium ion removal have the same structure, thus no volume change is observed during the charge-discharge process [23], which is important for the battery long term stability. Given the desired slim profile of the fiber-based battery (thicknesses of all the layers ~100μm), use of the zero-strain insertion materials becomes especially important. The choice for the anode material was therefore the spinel $Li_4Ti_5O_{10}$, which can accommodate three $Li^+$ ions per formula unit without any significant volume change during its transformation into the rock-salt $Li_4Ti_5O_{12}$ [24-27]. The discharging potential of this material is ~1.55 V vs. lithium metal, which is much higher than that for the graphite anodes. When combined with the $LiFePO_4$ cathode material a 1.8 V battery can be constructed. What is more important, the theoretical specific capacity of $Li_4Ti_5O_{12}$ is 175 mAh/g, which is well matched with that of $Li_4FePO_4$.

Another key material in a flexible battery, which is responsible for the battery unusual mechanical properties, is the solid polymer electrolyte (SPE). Within lithium batteries, polymer electrolyte plays two important roles. Firstly, it functions as an electron separator as well as an ion carrier between the highly reactive anode and cathode. Secondly, polymer electrolyte serves as a binder between cathode and anode. In addition to these two conventional uses of polymer electrolytes, for us, a very attractive feature of these materials is their thermo-elastic nature, which makes them suitable for extrusion and drawing techniques commonly used for fiber fabrication. Particularly, pure PEO can be successfully drawn above 70 C$^o$. However, the ionic conductivity of the PEO-based SPEs is high only at temperatures above the PEO melting temperature (~60 C$^o$), which narrows its practical application range. Since the discovery of ionic conductivity in a PEO/$Na^+$ complex in 1975, and the application of SPEs to lithium batteries [28], much effort has been made to improve the ionic conductivity of polymer electrolytes. The most investigated systems are the PEO-Li salt complexes, such as PEO-LiI, PEO-$LiCF_3SO_3$, PEO-$LiClO_4$ and PEO-$LiPF_6$. Additionally, some organic plasticizers or inorganic ceramic fillers, such as PEG, $TiO_2$, $Al_2O_3$, and $SiO_2$, are often added to improve the ionic conductivity of PEO at ambient temperatures [29, 30]. In our work we investigate the effect of various Li salts, as well as addition of the low molecular PEG on the ionic conductivity of PEO. Finally, we study the effect of the environmentally friendly aqueous solutions used in the battery preparation on the structure of polymer electrolytes.

## Experimental section

### Chemicals and materials

PEO (Mw = 400,000 g/mol), PEG (Mw=400 g/mol) were obtained from Scientific Polymer Products. Carbon black, LiI, $LiCF_3SO_3$, $LiPF_6$, $LiClO_4$ Cu and Al foil, PVDF, acetonitrile, ethylene carbonate (EC) and methylethyl carbonate (EMC) were obtained from Alfa Aesar. Electroactive $LiFePO_4$ and $Li_4Ti_5O_{10}$ were obtained from Phostech Lithium Co. Conducting Cu and Al wires are obtained from McMaster-Carr Supply Company. 100 % cotton threads are obtained from Coats & Clark Canada. All these materials were used as received without further purification.

### Samples Preparation

**Polymer electrolytes:**
Appropriate amounts of polymer and Li salt were first dissolved in aqueous (majority of experiments) or organic solvents (control experiments). These solutions were then poured either onto the glass substrate to cast a film or directly onto the anode or cathode films to make a multilayer film. The polymer electrolyte films were first dried in the hood under the horizontal air flow, followed by drying in the vacuum oven at 50 ºC.

### Electrode composites:
*Film electrode:*
The anode and cathode fabrication started with mixing the appropriate amounts of $Li_4Ti_5O_{10}$ or $LiFePO_4$ powder, PEO or PVDF powder (acting as binders), as well as electron conductive carbon black powder. The powder mix was then added into the PEO dissolved either in the aqueous or acetonitrile solution, and then mixed using magnetic stirrer. The resulting slurry was deposited onto a glass substrate, dried in the hood under the horizontal air flow, and then in the vacuum oven at 50 ºC (overnight) to get the anode and cathode films.
*Powder electrode:*
Powder electrode was prepared from the same powder mix as the film electrode. The mix was first pressed into a tablet, and then several drops of 10 % PEO solution or 5% PVDF solution were added on top of a tablet as a binder.

### Battery assembly:
In one approach, anode, polymer electrolyte and cathode films were first prepared separately and partially dried in the horizontal air flow. Then, all the layers were assembled, pressed to maintain a tight contact, and then dried at 50 ºC in the vacuum oven to obtain the final battery. In another approach, first, anode film was created and completely dried, then a solution for the separator layer was poured onto the anode layer and a two-layer system was created after drying. Finally, the cathode layer mix was poured onto the two layer system and dried to obtain the battery.

**Textile battery:**

Battery films were first cut into 1cm-wide ~10cm-long stripes. The battery strips were integrated into a textile during weaving with a manual Dobby loom. Cotton threads were holding the battery attached on the surface of a textile, while conductive threads were used to weave textile electrodes and to connect the individual battery stripes in series.

**Characterization**

WAXD was used to characterize the crystallinity of polymer electrolytes and the crystal structure of the electroactive materials. The WAXD measurements were carried out using a Bruker AXS diffractometer (Siemens Kristalloflex 780 generator) operated at 40 kV and 40 mA, using the Cu Kα (0.1542 nm) radiation collimated by a graphite monochromator and a 0.5 mm pinhole. The diffraction patterns were recorded by a HI-STAR area detector.

Electrical Conductivity. The conductivities of polymer electolytes were measured by electrochemical impedance spectroscopy using a potentiostat from Princeton Applied Research (model PARSTAT 2273). The test cell comprised two copper or aluminum electrodes with the area of ~1.26 cm$^2$. The thickness of the polymer electrolyte layers was measured using a caliper so that the conductivity could be obtained from the resistance.

Cyclic voltammetry (CV) was used to characterize the electrochemical activity of the electrode material. The cyclic voltammetry was measured with the same copper electrodes and the same potentiostat as in the electrical conductivity test.

Charge-discharge test was used to characterize the reversibility of the battery system. Cu and Al foils with the area of 1 cm$^2$ were used as electron conductors for cathode and anode films respectively. Constant current method (±0,02, ±0,05 or ±0,1 mA) was used in the test with the maximum charge or discharge time fixed at 0.5 hour. For the woven battery, charge-discharge characterisation is performed using 0.1mm-diameter Cu and Al wire electrodes woven at the time of sample preparation. The wires were held firmly at the appropriate faces of the battery stripes with the cotton threads.

# Results and discussion

### Effects of additives on the properties of polymer electrolyte

One of the key parameters affecting performance of a solid battery is the bulk electrolyte conductivity which characterizes ionic mobility in polymer electrolytes. The higher is the conductivity the more effective is the ion transfer across the battery. In a solid battery, the impedance between electrode/electrolyte interface, such as double layer capacitance $C_e$ as well as charge transfer resistance $R_e$, must be considered in addition to the bulk electrolyte resistance $R_s$. To understand battery performance, one typically assumes a certain effective electrical circuit of a battery such as the one shown in figure 1.

Detailed analysis of the equivalent circuit in figure 1 shows that complex part of the battery impedance will have two minima, one at lower frequencies with the corresponding value of the real part $Re(Z)=R_s+R_e$, and the other one at higher frequencies with the corresponding value of the real part $Re(Z)=R_s$. By measuring the bulk electrolyte resistance $R_s$ of a film sample and knowing the film thickness, one can extract the bulk electrolyte conductivity.

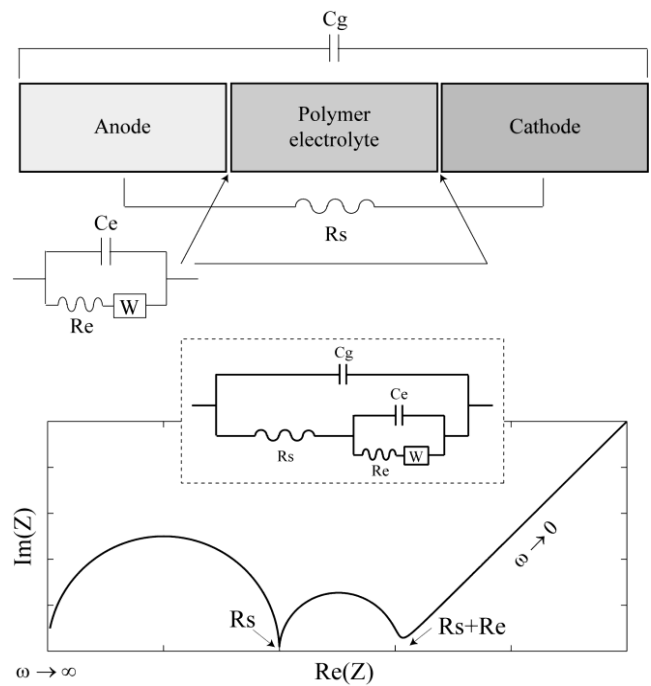

**Fig. 1**. The complex equivalent circuit for the battery system with polymer electrolytes. $C_g$ is the geometrical capacitance, $R_s$ is the polymer electrolyte resistance, $C_e$ is the electrode and electrolyte interfacial capacitance and $R_e$ is the electrode/electrolyte interfacial resistance, W is the Warburg impedance.

In what follows we present the ionic conductivities of PEO-LiX (X=I$^-$, CF$_3$SO$_3^-$, PF$_6^-$ and ClO$_4^-$) electrolytes measured with the ac impedance method described above. Two different electrode types were used. The first type included Cu or Al plates which are generally considered as lithium ion blocking electrodes. The second type included films cut from the cathode and anode sheets prepared from the LiFePO$_4$ and Li$_4$Ti$_5$O$_{10}$ materials. Results of our measurements are summarized in Table 1.

Firstly, we have investigated the effect of low molecular PEG (Mw = 400 g/mol) on the ionic conductivity of polymer electrolytes. As shown in Table 1, the values of the ionic conductivity measured are $1.54 \times 10^{-7}$, $2.28 \times 10^{-8}$ and $7.9 \times 10^{-9}$ Scm$^{-1}$ for PEG molar ratios of 50%, 25% and 10%, respectively. These values are all higher than for the pure PEO, which is $\sim 3.5 \times 10^{-9}$ Scm$^{-1}$. This indicates that addition of the low molecular weight PEG increases ionic conductivity of the polymer electrolyte which was also reported in [31].

Secondly, addition of Li salts (such as LiI or LiCF$_3$SO$_3$) into the polymer electrolytes increase dramatically the electrolyte ionic conductivity. Addition of the low molecular weight PEG further increases the ionic conductivity, however it has a much weaker influence on the conductivity when compared to the prior case without Li salts. The most important effect of PEG is however on the mechanical properties of the resultant films. Pure PEO films are highly crystalline and relatively rigid with a well defined melting temperature. Adding Li salts reduces crystallinity of PEO and for low concentration of salts the mix becomes soft and rubber like. At higher concentration of salts, however, the films lose their elasticity and start crumbling. Adding low molecular weight PEG into the PEO/Li salt combinations results in softer more elastic films even at high salt concentrations.

| Li salt | PEG ratio (y) | Urea ratio | Ionic conductivity |
|---|---|---|---|
| - | 0 | 0 | $3.50 \times 10^{-9}$ |
| - | 0.1 | 0 | $7.90 \times 10^{-9}$ |
| - | 0.25 | 0 | $2.28 \times 10^{-8}$ |
| - | 0.50 | 0 | $1.54 \times 10^{-7}$ |
| LiI | 0 | 0 | $1.67 \times 10^{-4}$ |
| LiI | 0.33 | 0 | $2.97 \times 10^{-4}$ |
| LiI | 0.50 | 0 | $9.23 \times 10^{-4}$ |
| LiI | 0.67 | 0 | $4.27 \times 10^{-4}$ |
| LiI | 0 | 0.4 | $2.16 \times 10^{-5}$ |
| LiI | 0 | 0.69 | $1.28 \times 10^{-5}$ |
| LiCF$_3$SO$_3$ | 0 | 0 | $2.05 \times 10^{-4}$ |
| LiCF$_3$SO$_3$ | 0.33 | 0 | $1.57 \times 10^{-4}$ |
| LiCF$_3$SO$_3$ | 0.50 | 0 | $3.33 \times 10^{-4}$ |
| LiCF$_3$SO$_3$ | 0.67 | 0 | $5.03 \times 10^{-4}$ |
| LiCF$_3$SO$_3$ | 0 | 0.4 | $1.22 \times 10^{-5}$ |
| LiCF$_3$SO$_3$ | 0 | 0.69 | $7.80 \times 10^{-6}$ |
| LiPF$_6$ | 0 | 0 | $3.88 \times 10^{-5}$ |
| LiPF$_6$ | 0 | 0.4 | $6.02 \times 10^{-5}$ |
| LiPF$_6$ | 0 | 0.69 | $1.60 \times 10^{-5}$ |
| LiClO$_4$ | 0 | 0 | $2.31 \times 10^{-5}$ |
| LiClO$_4$ | 0 | 0.4 | $4.11 \times 10^{-5}$ |
| LiClO$_4$ | 0 | 0.69 | $1.30 \times 10^{-5}$ |

Table 1. Ionic conductivity of (1-y)PEO-yPEG-LiX (X=I$^-$, CF$_3$SO$_3^-$, PF$_6^-$ and ClO$_4^-$) at room temperature. The molecular weight of PEO and PEG are 4,000,000 and 400 g/mol, respectively. The molar ratio of PEO(PEG):LiX is kept at 6:1 for all the samples, the urea molar ratios are 0.4 and 0.69 respectively, corresponding to the two complexes formed with PEO.

It has been known that Li salt can form complexes with PEO. The PEO chains are suggested to adopt a helical conformation with all C-O bonds *trans (t)* and C-C bonds either *gauche (g)* or *gauche minus (g-)*. Three ethylene oxide units are involved in the basic repeating sequence that is ttgttgttg¯. The Li$^+$ is located in each turn of the helix and is coordinated by the three ether oxygen in the case of Li salts [32-36]. Within the complexes, each cation is also coordinated by two anions and each anion bridges two neighbouring cations along the chain. Through our research, the ionic conductivity calculated based on the thickness and area of the electrolytes film are ~$2\times10^{-4}$ Scm$^{-1}$ for PEO-LiI and PEO-LiCF$_3$SO$_3$ films and ~$3\times10^{-5}$ Scm$^{-1}$ for PEO-LiPF$_6$ and PEO-LiClO$_4$ films at ambient temperature. This difference might come from the different anions in those Li salt and different degrees of crystallinity, which could be seen from the WAXD results presented later in the paper. For the practical use, the ionic conductivity should be above $1\times10^{-4}$ Scm$^{-1}$.

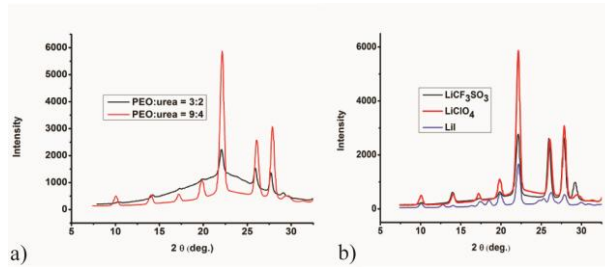

**Fig. 2** The WAXD results for (a) PEO-urea complexes with LiClO$_4$ and PEO:urea molar ratio of 3:2 and 4:9 (b) PEO- LiX (X= I$^-$, CF$_3$SO$_3$ and ClO$_4^-$)) with the PEO:urea molar ratio of 4:9.

The polymer electrolytes play three important roles in the battery. First, it is a lithium ion carrier; second, it is a separator between the two electrodes, which eliminates the need for an inert porous separator; third, it is a binder and an adhesive that ensures good mechanical and electrical contact with electrodes. As we have mentioned earlier, pure PEO films are highly crystalline and relatively rigid, while the ones with Li salts are more rubber like, especially the ones with low molecular weight PEG. The highly amorphous structures might facilitate ionic conductivity of the polymer electrolytes, and have a soft artificial leather-like feel, which is beneficial for the applications in wearables. At the same time, semi-crystalline structures with a controllable degree of crystallinity produces films with better mechanical properties and drawability. To control the degree of crystallinity we study adding the urea in the polymer composition. It has been reported that adding urea into the PEO film promotes crystallinity via formation of the highly crystallized complexes. Particularly, formation of specific complexes between PEO and urea was reported in [37,38] for the two PEO:urea molar ratios 3:2 and 4:9. The two complexes were suggested to be of a layered or channel type. As shown in Fig 2, the WAXDmeasurements of PEO-LiClO$_4$ compounds with high ratio of urea (PEO:urea=4:9) shows significant diffraction peaks, which means the high degree of crystallinity in the sample. Mechanically, these samples are brittle and disintegrate easily into pieces. When using the lower urea ratio (PEO:urea= 3:2), the WAXD measurements show both wide amorphous halo and sharp crystalline diffraction peaks. The crystalline peaks appear at virtually the same position for both the high and low urea ratios. This indicates that the crystalline structure of the complexes might be the same for both high and low ratios of urea. This phenomenon is the same for all the other Li salts tested in this work (see figure 2b). Overall we observe that adding urea promotes rigidity in the otherwise rubber-like films containing PEO-Li salt compositions, which can be highly beneficial for extrusion or drawing of these materials. Finally, in Table 1 we present the ionic conductivity of compounds containing different molar ratios of urea in the PEO. We find that for PEO-LiClO$_4$ and PEO-LiPF$_6$ compounds, the ionic conductivities are comparable to each other with or without urea. However, for PEO-LiI and PEO-LiCF$_3$SO$_3$ compounds ionic conductivity drops by an order of magnitude when urea is added. In all the cases, of the two samples with different ratios of urea, the one with smaller urea content (samlpes of lower crystallinity) has consistently higher conductivity that the one with higher urea content (samples of higher crystallinity).

**Effects of additives on the properties of electrodes**

A battery electrode has to exhibit simultaneously good electron and ionic conductivities. In the case of a cathode, for example, pure $LiFePO_4$ exhibits low electron conductivity, thus, electron
5 conductors have to be added into a cathode compound. In fact, in a standard battery, to form the electrodes one typically uses powder compositions of various electroactive materials mixed with small amounts of a binder. The electrode pallets are then created by forming the powder mix under press. In our case, the
10 goal is to create extrudable/drawable electrodes, therefore, a larger quantity of polymer binder materials has to be used in order to obtain the desired thermo-mechanical properties of the electrode material. In figure 3 we present examples of electrodes and battery samples prepared by solution casting method using
15 PEO as a binder and carbon black as electron conductive material. The cathode, anode, polymer electrolytes and complete batteries are all soft and highly stretchable; moreover, they have a feel and appearance of artificial leather, which is highly appropriate for applications in wearables. The 1cm x 10cm
20 battery stripes cut from the planar film samples have very robust mechanical properties, and can be easily weaved into textiles.

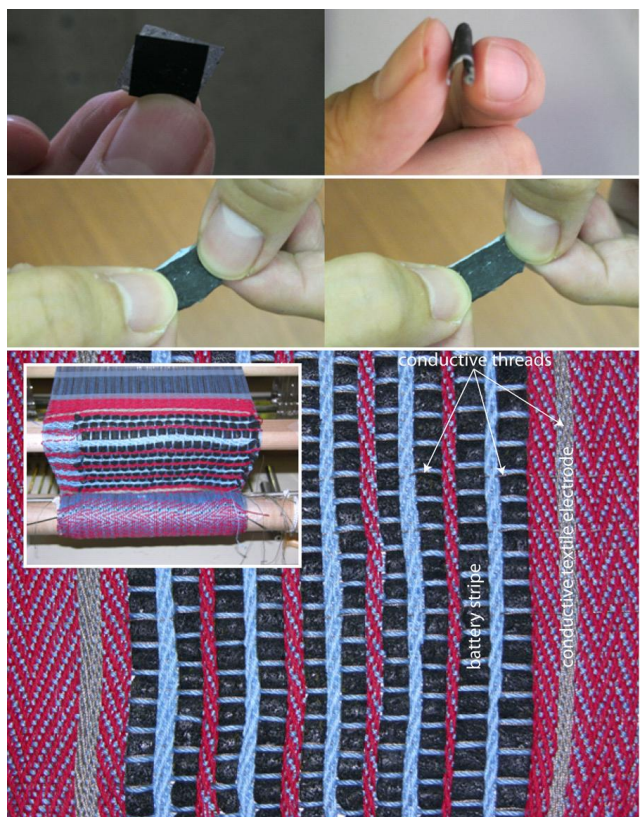

**Fig. 3** Top row: photographs of a flexible battery made of binding individual cathode, anode and polymer electrolyte films. Middle row:
25 resulting battery is highly stretchable. Bottom row: battery stripes (black) woven into a textile (blue and red cotton threads) using Dobby loom . The stripes are connectorized in series with conductive threads (metallic brown). Two textile electrodes are formed by the conductive threads at the textile extremities.

30 The electron and ionic conductivities have been measured with the dc and ac methods respectively. The electronic conductivities of both the cathode and anode were $\sim 1\times 10^{-4}$ S cm$^{-1}$, which is much higher than those of pure PEO, $LiFePO_4$ or $Li_4Ti_5O_{10}$ powders. However, compared to the conventional electron
35 collecting materials, such as copper or aluminum, the electronic conductivity of the soft electrodes is still very small.

To investigate the effect of PEO ratio on the properties of electrodes, two types of samples were prepared. The first series of
40 samples has low PEO content (less than 5%) where PEO acts mainly as a binder material to hold the powder together as in the conventional Li battery. In particular, the powder cathode and anode are composed of 87% $LiFePO_4$ or $Li_4Ti_5O_{10}$ and 13% carbon black, then binded with a few drops of 5% PEO solutions.
45 The second series of samples has high PEO content (above 25 %) and the resultant electrodes are flexible films. In these samples the cathode and anode films are composed of 37.5% $LiFePO_4$ or $Li_4Ti_5O_{10}$, 50% PEO and 12.5% carbon black. In Fig. 4 we present a typical result of the cyclic voltammetry measurements.
50 For example, an anode made of pressed powder exhibits an oxidation current peak which is much larger than that of an anode film with high PEO content. A similar effect is observed for the powder and film cathodes. While voltammetry results indicate that large resistance is indeed brought by the high PEO content, at
55 the same time they also show that reversibility of a film battery is at least as good as the reversibility of the powder-based battery. This is judged from the good repeatability of the I(V) curves during 5 cycles of the voltammetry experiment.

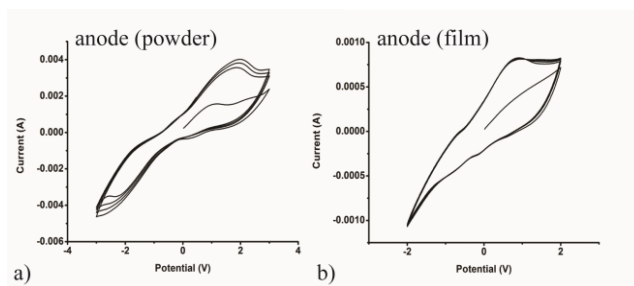

60 **Fig. 4** The cyclic voltammetry results of a) anode powder sample, b) anode film sample.

**Electrochemical properties of flexible batteries**

**Open circuit voltage measurements:**
65 In this section we report performance of several batteries assembled with various material choices for anode, cathode and electrolyte. Based on our measurements, we conclude that $LiFePO_4$, $Li_4Ti_5O_{10}$, PEO material composition presents a viable flexible all-solid battery system, however, the registered voltage
70 is always significantly lower than the theoretical value of 1.8 V.

All the batteries in our experiments can be characterised as those with powder pressed electrodes (no or little PEO) or film electrodes (high ratio of PEO binder). Moreover, in our
75 experiments we compare battery performance when using solid electrolyte separator layer versus a filtration paper soaked in liquid electrolyte. Electrode and electrolyte types and compositions, as well as open circuit voltage (OCV) of the corresponding batteries are listed in table 2.
80

First, we have tested performance of a battery comprising powder anode and cathode reported in the previous section, while using as a separation layer a filtration paper soaked either in PEO(PEG):LiI aqueous solution or PEO:LiPF$_6$ in the EC/EMC (1:1) solution. Not surprisingly, batteries comprising powder electrodes and liquid electrolyte showed consistently the best performance with the highest open circuit voltage ~1 V.

In the next set of experiments we have retained a filtration paper soaked in liquid electrolyte as a separator layer, while substituting powder pressed anode and cathode with film anode and cathode described in the previous section. Two types of liquid electrolited were tested including PEO:LiPF$_6$ in the EC/EMC (1:1) solution and PEO:LiCF$_3$SO$_3$ in acetonitrile solution. In both systems, OCV dropped from ~1V to ~0.7V. This result correlates with the greatly reduced ionic and electronic conductivity of the PEO containing electrodes compared to the powder pressed electrodes.

| Li salts and electrolyte types | PEG ratio (y) | Urea ratio | Electrode Types | OCV (V) |
|---|---|---|---|---|
| LiI solution | 0.50 | 0 | powder | 1.00 |
| LiPF$_6$ solution | 0 | 0 | powder | 1.00 |
| LiPF$_6$ solution | 0 | 0 | film | 0.72 |
| LiCF$_3$SO$_3$ solution | 0 | 0 | film | 0.70 |
| LiCF$_3$SO$_3$ film | 0 | 0 | powder | 0.50 |
| LiI film | 0 | 0 | powder | 0.32 |
| LiI film | 0.33 | 0 | powder | 0.36 |
| LiI film | 0.50 | 0 | powder | 0.52 |
| LiI film | 0.67 | 0 | powder | 0.56 |
| LiI film | 0 | 0.40 | powder | 0.63 |
| LiI film | 0 | 0.69 | powder | 0.52 |
| **LiI film** | **0** | **0.69** | **film** | **0.50** |
| **LiI film** | **0.50** | **0** | **film** | **0.45** |

Table 2. Open circuit voltage measured with various electrolytes (polymer solution and polymer solid) and two types of electrodes (powder electrode and film electrode). LiI solution refers to aqueous solution, LiPF$_6$ solution refers to the ethylene carbonate (EC) / ethylmethyl carbonate (EMC) (1:1) solution, LiCF$_3$SO$_3$ solution refers to the acetonitrile solution. The molar ratio of PEO$_{1-y}$(PEG$_y$): Li-X is kept at 6:1 for all the compositions.

Most pronounced effect on the OCV was observed when we have substituted liquid electrolyte-based separator layer with solid electrolyte film. In what follows, all the solid electrolyte films had the composition PEO$_{1-y}$(PEG$_y$):Li-X, where a constant 6:1 molar ratio was used for the polymer to salt ratio. In the first set of experiments we have retained powder anode and cathode and used solid electrolyte films only as a separator layer. When using PEO:LiCF$_3$SO$_3$ electrolyte the OCV dropped to 0.5V, however the reduction was worth in the case of a PEO:LiI electrolyte for which the OCV was ~0.32V. By adding significant amounts of the low molecular weight PEG or urea into PEO:LiI electrolyte it was possible to increase the OCV to ~0.5-0.6V. These results correlate perfectly with the ionic conductivity measurements presented in Table 1. Namely, higher OCV values are consistently achieved in systems with higher ionic conductivities of the solid polymer electrolytes used in a separator layer.

Finally, when substituting the powder pressed anode and cathode with their film homologues, no significant voltage drop was observed. This allowed us to obtain OCV of ~0.5V in the all-solid battery systems comprised of solid electrodes separated with PEO:LiI electrolyte films that ether contained high ratios of low molecular weight PEG or urea. Although in both cases battery structure was rubber-like with mechanical properties mostly determined by the soft outer electrodes, urea containing batteries were tangibly firmer than those containing PEG.

**Charge-discharge measurements:**

Although an open circuit voltage is an important indicator of the battery performance, the more important test is a charge-discharge cycling under loading. In figure 5 we present constant current (±0.02mA, ±0.05mA and ±0.1mA) charge-discharge tests of the two 1cm x 1cm battery samples, each containing the same PEO:LiI (6:1) polymer electrolyte separator layer. Copper and aluminium foils were used as electron collectors in the measuring cell. The film electrodes in the first battery sample (figure 5(a)) were prepared using PEO (26.7%), LiFePO4 or Li$_4$Ti$_5$O$_{10}$ (66.7%), and carbon black (6.6%) (by weight). The second sample featured film electrodes with higher concentration of PEO and carbon black, namely, PEO (50%), LiFePO4 or Li$_4$Ti$_5$O$_{10}$ (37.5%), carbon black (12.5%). For the first sample with electrodes containing smaller amounts of PEO (see figure 5(a)), the discharge curves at currents 0.02mA and 0.05mA showed a continuous decay from ~0.5V to ~0.1V with no change in the discharge time after 5 cycles. At higher currents (0.1mA) discharge time somewhat shortened during the first five discharges with the discharge voltage dropping to zero after 3 cycles. For the second sample with electrodes containing larger amounts of PEO (see figure 5(b)), at currents 0.02mA and 0.05mA the discharge curves first show an almost instantaneous drop from 0.4-0.5V to ~0.3V followed by a slow linear in time decay. No change in the discharge time is observed after 5 cycles at lower currents. At higher currents (0.1mA) discharge time shortened significantly during the first five discharges with the discharge voltage dropping to zero already after the first cycle.

These and similar charge-discharge experiments consistently show that initial discharge voltage is higher and it decreases slower in battery samples featuring electrodes with lower amounts of PEO. At the same time, it appears that battery samples containing electrodes with higher amounts of PEO show a better performance at longer discharge times, where voltage decrease is relatively slow and almost linear with time. Overall, these charge-discharge experiments indicate good reversibility of the solid electrolyte-based batteries developed in this work even though a typical measured discharge voltage ~0.2-0.3V is much lower than the theoretical one of 1.8V.

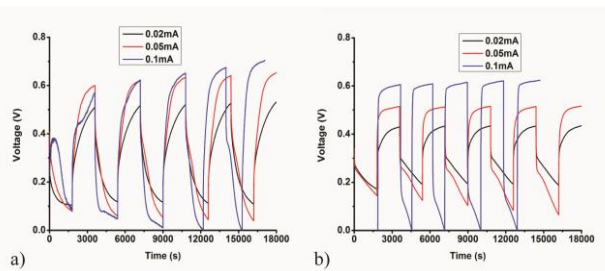

**Fig. 5** Constant current charge-discharge curves of the two flexible batteries with a solid PEO:LiI polymer electrolyte separator layer. a) Electrodes with 26.7% of PEO. b) Electrodes with 50 % of PEO.

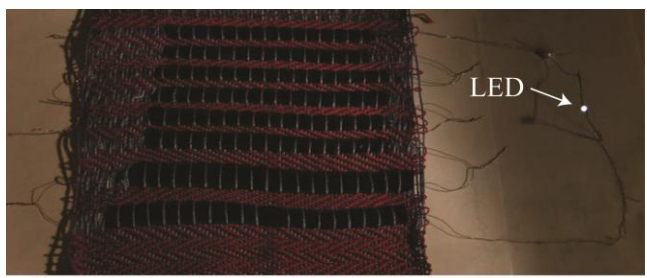

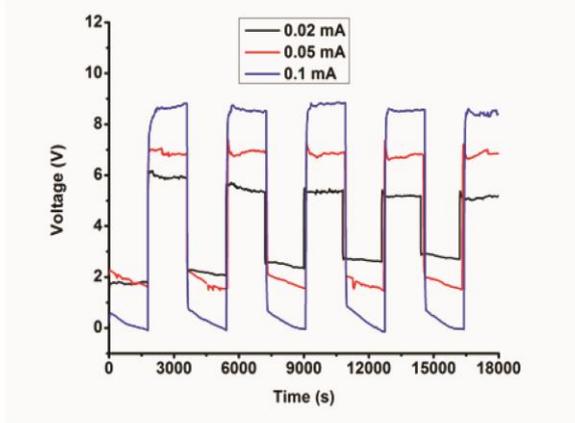

**Fig. 6** Top: Textile battery is made of 8 battery stripes woven with cotton thread and connectorized in series using copper and aluminium wires (one per stripe per side) as electron collectors. The resultant battery is powerful enough to light up a 3V LED for several hours. Bottom: the charge-discharge curves of the textile battery.

Although operating voltage of a single flexible battery is relatively low (~0.3V), when several of them are connected in series, the net voltage can be large enough for practical applications. In figure 6 we present an example of a textile battery comprising 8 flexible battery stripes woven together and connectorized in series to power up a 3V LED. This battery provides dim LED light for several hours and it can be recharged. The electrode compositions used in this sample are those described above with high content of PEO (50%). Charge-discharge curves for the textile battery were measured after connectorization of all the stripes in series using copper and aluminium wires (one wire per stripe per side). The charge-discharge curves showed stable discharging plateaus at ~2V for lower currents of 0.02 mA and 0.05 mA, while at higher current of 0.1mA the discharging voltage rapidly dropped to zero. At the same, very high values of the charging voltages 5.5, 7 and 8.5 V for the charging currents of 0.02, 0.05 and 0.1 mA indicate that the internal resistance of a textile battery is high. This is in part due to a relatively small contact area between the battery polymer electrode and the electron collector in the form of thin wires. Note that, in principle, charging voltages can be reduced by using metallic foils with large surface area as electron collectors instead of wires. However, in textile applications the most appropriate is to use wires or conductive threads as the electron collectors (see figure 3), as they can be naturally integrated during weaving.

**Effect of solvents and PEO on electrode structure:**

As seen from the table 2, open circuit voltages of all the film batteries are much lower than the theoretical value (1.8 V). This can be attributed to the changes in the physical and chemical structure of the electrodes after treatment of the pure powders of $LiFePO_4$ or $Li_4Ti_5O_{10}$ with solvents and addition of PEO. As a result, the electrochemical reaction at the interfaces between electrodes and electrolyte might change. Here we use WAXD to probe differences in the structure of pure $LiFePO_4$ and $Li_4Ti_5O_{10}$ powder electrodes versus film electrodes that contain significant amounts of PEO and were subject to solvent treatment.

In figure 7(a) we present WAXD results for cathode. Particularly, we compare diffraction peaks coming from the pure $LiFePO_4$ powder electrode to the diffraction peaks coming from the film electrode containing 37.5% PEO, 50% $LiFePO_4$, 12.5% carbon black and cast from the aqueous solution of PEO. For comparison, WAXD of a pure PEO powder sample is presented on the same plot. All the diffraction peaks of $LiFePO_4$ could be indexed with an orthorhombic structure (a= 10.323Å, b= 6.003 Å and c=4.694 Å) [39, 40]. From figure 7(a) we see that all the peaks in the film cathode can be related to the peaks of pure $LiFePO_4$ or PEO materials, which means that the chemical structure of a cathode film is similar to that of the basic elements used in its fabrication. Physical structure of the cathode is clearly semi-crystalline as judged from the broad and relatively intense background.

In figure 7(b) we present WAXD results for anode. There, diffraction peaks coming from the pure $Li_4Ti_5O_{10}$ powder electrode are compared to the diffraction peaks coming from the film electrode containing 37.5% PEO, 50% $Li_4Ti_5O_{10}$, 12.5% carbon black and cast from the aqueous solution of PEO. All the diffraction peaks of $Li_4Ti_5O_{10}$ could be indexed with a cubic spinel structure (a=b=c=8.376 Å) [41]. From figure 7(a) we see that diffraction peaks corresponding to the anode film are quite different from those corresponding to the powder anode. For example, the most intense band at ~37 ° in the powder sample is missing in the film sample. Difference in the chemical and physical structure of an anode material after its treatment with aqueous solution of PEO can be one of the reasons why measured open circuit voltage is different from the theoretical prediction. Additionally in figure 7(b) we present WAXD results for a $Li_4Ti_5O_{10}$ powder sample treated with acetonitrile solution, and observe no change in the diffraction bands of an anode material. Finally we note that anode film shows high degree of crystallinity as judged from the low intensity of the broad background.

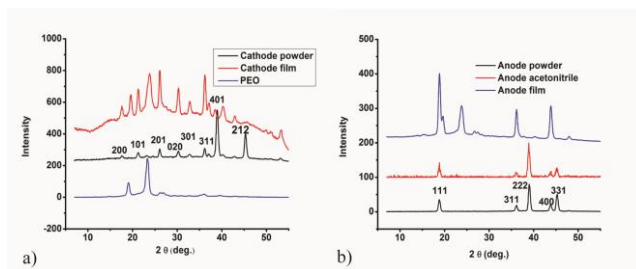

**Fig.7** WAXD results for the a) powder (no PEO) and film (50% PEO) cathode b) powder (no PEO) and film (50% PEO) anode.

## Conclusions

Flexible and stretchable film batteries for smart textile applications have been demonstrated with conventional Li battery materials including LiFePO$_4$ cathode, Li$_4$Ti$_5$O$_{10}$ anode and PEO solid electrolyte. By introducing large quantities of the thermoplastic PEO binder in the battery electrodes and separator layer one can potentially realise a fully extrudable/drawable battery system, which could allow direct drawing of battery fibers ideal for textile applications. Alternatively, we have experimentally demonstrated that flexible batteries can be first cast as sheets, then cut into thin stripes, and finally integrated into textile using conventional weaving techniques. The electrochemical performance of the film batteries was extensively characterised and found to be poorer compared to the performance of batteries based on the powder electrodes and liquid electrolytes. At the same time, cycling performance of the solid film batteries was stable, and together with their soft leather-like feel and appearance, this makes such batteries well suitable for smart textile applications. Finally, the film batteries were made using environmentally friendly fabrication route, where in place of organic solvents only aqueous solutions were used to cast the electrodes and solid electrolyte separator film.

## Acknowledgement

The authors would like to thank Prof. L. Martinu and Dr. J. Sapiena at Ecole Polytechnique de Montréal for their help in electrical characterization of the samples. We would also like to thank Prof. D. Rochefort at University of Montréal for his help in ionic conductivity characterization. Finally, we thank Phostech Lithium Co. for providing cathode and anode materials.

## Notes and references

*a* Ecole Polytechnique de Montréal,Génie Physique, C.P. 6079, succ. Centre-ville, Montréal (Québec), Canada  H3C 3A7, Tel: (514) 340-4711 (3327), Fax: (514) 340-3218